\documentclass[conference]{IEEEtran}
\usepackage{cite}
\usepackage{amsmath,amssymb,amsfonts}
\usepackage{algorithmic}
\usepackage{graphicx}
\usepackage{textcomp}
\usepackage{xcolor}
\def\BibTeX{{\rm B\kern-.05em{\sc i\kern-.025em b}\kern-.08em
    T\kern-.1667em\lower.7ex\hbox{E}\kern-.125emX}}

\usepackage{bm}
\usepackage[inline]{enumitem}
\usepackage{array}
\usepackage{tabularx}
\usepackage{longtable, tabu}

\usepackage[outdir=./]{epstopdf}

\definecolor{light-gray}{gray}{0.8}

\begin{document}

\title{Cluster Lifecycle Analysis: Challenges, Techniques, and Framework}

\author{
    \IEEEauthorblockN{Ivens Portugal}
    \IEEEauthorblockA{\textit{David R. Cheriton} \\
        \textit{School of Computer Science} \\
        \textit{University of Waterloo} \\
        Waterloo, Canada \\
        iportugal@uwaterloo.ca}
    \and
    \IEEEauthorblockN{Paulo Alencar}
    \IEEEauthorblockA{\textit{David R. Cheriton} \\
        \textit{School of Computer Science} \\
        \textit{University of Waterloo} \\
        Waterloo, Canada \\
        palencar@cs.uwaterloo.ca}
    \and
    \IEEEauthorblockN{Donald Cowan}
    \IEEEauthorblockA{\textit{David R. Cheriton} \\
        \textit{School of Computer Science} \\
        \textit{University of Waterloo} \\
        Waterloo, Canada \\
        dcowan@csg.uwaterloo.ca}
}

\maketitle

\begin{abstract}
Novel forms of data analysis methods have emerged as a significant research direction in the transportation domain. These methods can potentially help to improve our understanding of the dynamic flows of vehicles, people, and goods. Understanding these dynamics has economic and social consequences, which can improve the quality of life locally or worldwide. Aiming at this objective, a significant amount of research has focused on clustering moving objects to address problems in many domains, including the transportation, health, and environment. However, previous research has not investigated the lifecycle of a cluster, including cluster genesis, existence, and disappearance. The representation and analysis of cluster lifecycles can create novel avenues for research, result in new insights for analyses, and allow unique forms of prediction. This technical report focuses on studying the lifecycle of clusters by investigating the relations that a cluster has with moving elements and other clusters. This technical report also proposes a big data framework that manages the identification and processing of a cluster lifecycle. The ongoing research approach will lead to new ways to perform cluster analysis and advance the state of the art by leading to new insights related to cluster lifecycle. These results can have a significant impact on transport industry data science applications in a wide variety of areas, including congestion management, resource optimization, and hotspot management.
\end{abstract}


\section{Introduction}

Novel forms of data analysis methods have emerged as a significant research direction in the transportation domain. As one of the major themes, research in transportation data science aims at understanding the movement of people, vehicles, or goods in the space over time \cite{Fischer199581}. The current literature includes research works in prediction \cite{Ma2015, Ma20181135}, optimization \cite{Rios-Mercado2015536, Wang2018925}, modeling \cite{Fischer199581, Huang2018251}, and analysis \cite{Li2018213, Satyakrishna2018268}, among others. Existing research has addressed several significant problems, including the scheduling of shipping jobs \cite{Fischer199581}, congestion management \cite{Zarmehri201613}, and the optimization of the use of resources \cite{Khazaei2016419}.

Transportation data analysis has direct economic and social implications on an organization either by improving the dynamics of the movement of elements, or by avoiding negative outcomes. For example, predicting traffic congestion helps users of a mobile app to select a better route to reach their destination, thus saving time; the analysis of customer movement in a supermarket helps managers in decision-making, thus increasing profits. When transportation data analysis is overlooked, the negative consequences may affect the profits, the environment, the well-being of people, and may make planning more difficult. According to \cite{Fischer199581}, some decades ago, when transportation data analysis was not performed, almost 40\% of truck in Europe drove empty. This has negative consequences to the economy and on the environment of the continent.

Specifically, clustering techniques are widely used in transportation data analysis. Clustering is the task of grouping elements based on their similarity, forming clusters, that can be used for classification or outlier detection. In the transportation domain, clustering techniques are used for many applications, including public transportation analysis \cite{Andrade201413}, transportation infrastructure improvements \cite{Rajput2017129}, and logistics \cite{Wang2016}.

Although the use of clustering techniques in transportation analysis has recently improved, its impact remains limited. There is not a clear understanding of the formation, existence, and disappearance of clusters when analyzing moving elements, such as people, vehicles, or goods. How does a cluster of tourists visiting a museum behave? And how do they differ from a cluster of students in a school trip? Why does a cluster of taxis form? What are the implications on the demand of taxis? How to analyze the movement pattern of a cluster of goods in one ship that moves to other different cluster of goods in other ships? Answers to these questions allow novel perspectives on the data and create new avenues for research.

The big data revolution of the last decade changed the way data analysis is done in several domains, including the transportation as well \cite{Torre-Bastida2018742}. The widespread use of GPS devices resulted in a massive generation of spatial-temporal data, which is the data described by its location and time of the measurement. Big data is usually described in terms of its four Vs, namely the volume, variety, velocity, and value \cite{Berman2013}. The transportation domain is a big data area because of the amount of data generated by GPS devices (volume), the many different types of transportation means (variety), the dynamic nature of the data and the near real time need for analysis results (velocity), and the economic and social implications of its analysis (value) \cite{Zheng2016620}. The importance of big data frameworks for transportation data analysis is clear. However, there is a lack of bid data frameworks addressing the understanding of the formation, existence, and disappearance of clusters of moving elements.

This study addresses two problems: \begin{enumerate*}[label=(\roman*)] \item the lack of a cluster lifecycle analysis (i.e. cluster genesis, existence, and disappearance) in the transportation domain, and \item the lack of a framework to perform such analysis \end{enumerate*}. This ongoing research proposes a study on the relations between clusters and moving elements or other clusters, their impacts on the cluster formation, existence, and disappearance, and a cluster lifecycle from start to end of its lifetime. Moreover, this research proposes a big data framework that helps in the processing of transportation data, analysis, and storage.

This technical report is structured as follows. Section \ref{application} discusses, as motivation, transportation application scenarios that can benefit from cluster lifecycle analysis. Section \ref{approach} describes the proposed study and framework. Section \ref{conclusion} presents conclusions and discusses some future work opportunities.

\section{Application Scenarios}
\label{application}

Cluster lifecycle analysis represents a novel perspective in how research on the cluster of moving elements can be done. There are many questions that can be addressed by this research. Table \ref{questions} presents some type of questions that cluster lifecycle analysis tries to understand. The following sections discuss how these questions relate to three transportation problems, namely the traffic congestion management, the optimization of resources, and hotspot management.

\begin{table*}
\caption{Types of questions for transportation applications.}
\begin{tabular}{p{5.0cm}p{12.0cm}}
\hline
Types                                   & Questions                                                                                                                        \\\hline
Cluster behavior                        & How does a cluster behave over time?                                                                                             \\
Cluster lifecycle similarity            & How different are a cluster of tourists visiting a museum and a cluster of students in a school trip in terms of their behavior? \\
Cluster formation                       & How are clusters formed? Which clusters merge?                                                                                   \\
Cluster size                            & What are the sizes of the formed clusters?                                                                                       \\
Cluster lifetime                        & How long do clusters exist?                                                                                                      \\
Cluster resource supply and demand      & How does cluster lifecycle relate to resource demand and supply?                                                                 \\
Cluster element dynamics                & How do elements enter of leave clusters over time through the lifecycle?                                                         \\
Cluster element persistence             & How do elements persist or remain in the cluster?                                                                                 \\
Cluster disappearance                   & How do clusters disappear? Do they split into different clusters?                                                                \\
Cluster formation and disappearance rates & What is the rate of cluster formation and disappearance?          \\\hline
\end{tabular}
\label{questions}
\end{table*}

\subsection{Traffic Congestion Management}

Traffic congestion is both a physical phenomenon and one that relates to the user experience, and therefore, a clear definition is difficult to provide \cite{NoAuthor20071}. It relates to both the traffic that nears the capacity of the road system, and the possible gap between the user's expectations of the traffic and the real traffic \cite{NoAuthor20071}. In any definition, traffic congestion includes a large cluster of elements that move slowly for some time. A study on how these clusters are formed, their duration, and how they disappear can lead to novel knowledge for traffic congestion understanding and prediction. Is it possible to predict a traffic congestion based on the movement (e.g. a cluster ``approaching'' a specific place) of cluster of vehicles in a city? How long do traffic congestion points last? What causes them to disappear and how do they disappear (e.g. split into smaller clusters)?

\subsection{Optimization of Resources and Logistics}

Logistics is the task of coordinating people, vehicles, or goods in the space \cite{Lun20101}. This task is challenging because of the many constraints that it imposes. For example, when shipping a container of industrialized products, the container moves on air, sea, and ground, together with other containers, until its final destination. The booking of an airplane ticket is a similar situation. One passenger may take many different routes in his or her trip, with other passengers. The study of how elements move between clusters, or how clusters of elements move between clusters can provide new insights. How many passengers remain on the same route after a long trip with many stops? How similar are these clusters? Is it possible to follow a small cluster while it joins and leaves other clusters?

\subsection{Hotspot Management}

A hotspot is a place of interest for some group of people. It may be a music concert or a touristic place. The essential concept is that it attracts people for some time \cite{Woodside20071}. Some of these events are predictable (e.g. a sports match) and others are spontaneous (e.g. a street performance). The detection or prediction of these hotspots have great implications on transporting elements. For example, a taxi driver may avoid a street close to a stadium to avoid being stuck in a traffic jam, while another taxi driver may seek to profit by taking spectators to and from the stadium. What is the size of cluster of spectators? How do they differ depending on the event type? Is it possible to predict the end of a cluster, so that elements will start moving away from it at a faster rate?

\section{Trajectory Cluster Lifecycle Analysis}
\label{approach}

\subsection{Overview}

A study of the formation, existence, and disappearance of clusters directs research to novel conclusions about cluster dynamics. To perform such studies, many trajectories are gathered and analyzed, and the results are used to identify clusters. These clusters are observed from conception to disappearance, including their relationships with other trajectories or clusters. Finally, a cluster lifecycle can be extracted from these observations, which supports new forms of analysis.

Specifically, a trajectory dataset is processed for cluster identification. Since each trajectory has timestamps taken at different rates, the study assumes a universal timestamp and queries each trajectory for the data at its next valid timestamp. A clustering algorithm, DBSCAN \cite{Ester:1996:DAD:3001460.3001507}, is run at the universal timestamps to detect clusters. At every timestamp, calculations are made to detect cluster similarity and behavior such as trajectories entering a cluster or a cluster splitting into two clusters based on the number of elements on each cluster at each timestamp. Results are saved to form a lifecycle for each cluster.

\subsection{Relations and Algorithm}

Prior to a discusion on the relations between clusters and trajectories or other clusters, this technical report describes some formal definitions used to guide this study. Table \ref{definitions} presents a list of such definitions.

\begin{table}
\caption{Formal Definitions.}
\tabulinesep=1mm
\begin{tabu} to \linewidth {| X[3.5,c,m] | X[5.0,j] |}

    \hline
    Definition & Explanation \\
    \hline

    $p_{i,t_k}$ & \begin{minipage}{\hsize}Point of trajectory $i$ at timestamp $t_k$.\end{minipage} \\ \taburulecolor{light-gray} \hline \taburulecolor{black}

    $p_{i,t_{i,k,stop}}$ & \begin{minipage}{\hsize}Point of trajectory $i$ at the moment trajectory $i$ started its most recent stop, based on $t_k$.\end{minipage} \\ \taburulecolor{light-gray} \hline \taburulecolor{black}

    $c_{i,t_k}$ & \begin{minipage}{\hsize}Center of group $g_i$ at timestamp $t_k$.\end{minipage} \\ \taburulecolor{light-gray} \hline \taburulecolor{black}

    $c_{i, t_{i,k,stop}}$ & \begin{minipage}{\hsize}Center of group $i$ at the moment it started its most recent stop, based on $t_k$.\end{minipage} \\ \taburulecolor{light-gray} \hline \taburulecolor{black}

    $dist(p_{i,t_k}, p_{j,t_k})$ & \begin{minipage}{\hsize}The distance between trajectories $i$ and $j$ at timestamp $t_k$. Usually euclidean distance.\end{minipage} \\ \taburulecolor{light-gray} \hline \taburulecolor{black}

    $movement(p_{i,t_k})$ & \begin{minipage}{\hsize}The movement type in which trajectory $i$ is at timestamp $t_k$.\end{minipage} \\ \taburulecolor{light-gray} \hline \taburulecolor{black}

    $group(c_{i,t_k})$ & \begin{minipage}{\hsize}The group whose with center $i$ at timestamp $t_k$.\end{minipage} \\ \taburulecolor{light-gray} \hline \taburulecolor{black}

    $belong(p_{i,t_k}, group(c_{i,t_k}))$ & \begin{minipage}{\hsize}$True$ if the point $p_{i,t_k}$ belongs to the group whose center is $c_{i,t_k}$ at timestamp $t_k$ according to the cluster algorithm being used. False otherwise.\end{minipage} \\

    \hline
\end{tabu}
\label{definitions}
\end{table}

\begin{figure}
    \centering
    \includegraphics[width=3in]{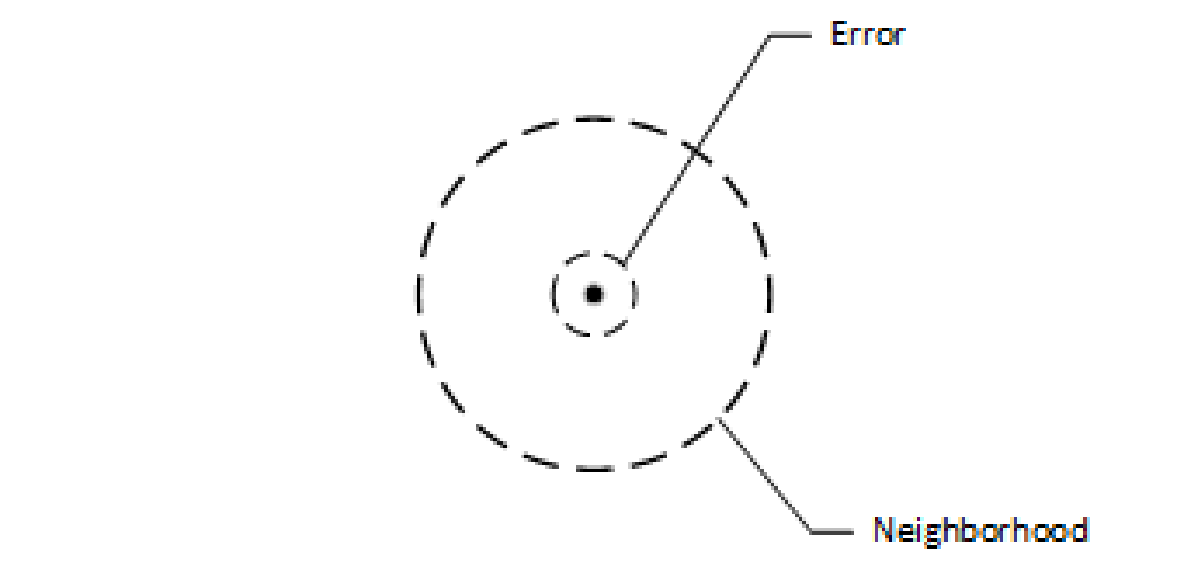}
    \caption{The parts of a trajectory.}
    \label{trajectory}
\end{figure}

A moving element, or trajectory, represents an element whose location may change (e.g. latitude, longitude) over time. The location of the element may also be the same over time to represent stops. As can be seen in Figure \ref{trajectory}, a trajectory is composed of two regions. The first one is an \textit{error} threshold. It accounts for errors during data capture. The second one is the \textit{neighborhood} threshold. It accounts for relationships that are near the element whose trajectory is being analyzed, but not at the same location.

Formally, a trajectory $traj_i$ is a series of points $p_{i,t_k}$, such that $traj_i = [p_{i,t_1}, p_{i,t_2}, ..., p_{i,t_n}]$, where $n$ is the length of the trajectory. Each $p_{i,t_k}$  has two coordinates $x$, $y$ (e.g. latitude and longitude), and a timestamp $t$, such that $p_{i,t_k} = (x_i(t_k), y_i(t_k), t_k)$. The first trajectory region, the \textit{error} threshold, is defined based on the radius $r_e$, whereas the second region, the \textit{neighborhood} threshold, is defined based on the radius $r_n$. The regions can be seen as the same at all points of the existence of each trajectory, but this is not necessarily true. Note that $t_{i,k,stop}$ refers to the timestamp where $traj_i$ started its most recent stop, based on a given $t_k$, and $p_{i,t_{i,k,stop}}$ refers to that point. To formalize $t_{i,k,stop}$, let $P_i=\{p_{i,t_\alpha},p_{i,t_\beta},...,p_{i,t_\omega}\}$ be the set of all points of $traj_i$, of lenth $n$, $t_1 \leq t_\alpha, t_\beta, ..., t_\omega \leq t_n$, where $movement(p_{i,t_{k-1}})=\textrm{\textit{move}}$ and $dist(p_{i,t_k}, p_{i,t_{k-1}})<r_e$. Let $T_i=\{t_\alpha, t_\beta, ..., t_\omega\}$ be the set of all timestamps of the points in $P_i$. At any given timestamp $t_k$, $t_{i,k,stop}=\min_{t_j\in T_i}(t_k-t_j)$, $t_j < t_k$.

In summary, trajectories have two main types of movements: \textit{stop} and \textit{move}. They are graphically represented and formalized in Table \ref{movements}.

\begin{figure}
    \centering
    \includegraphics[width=3in]{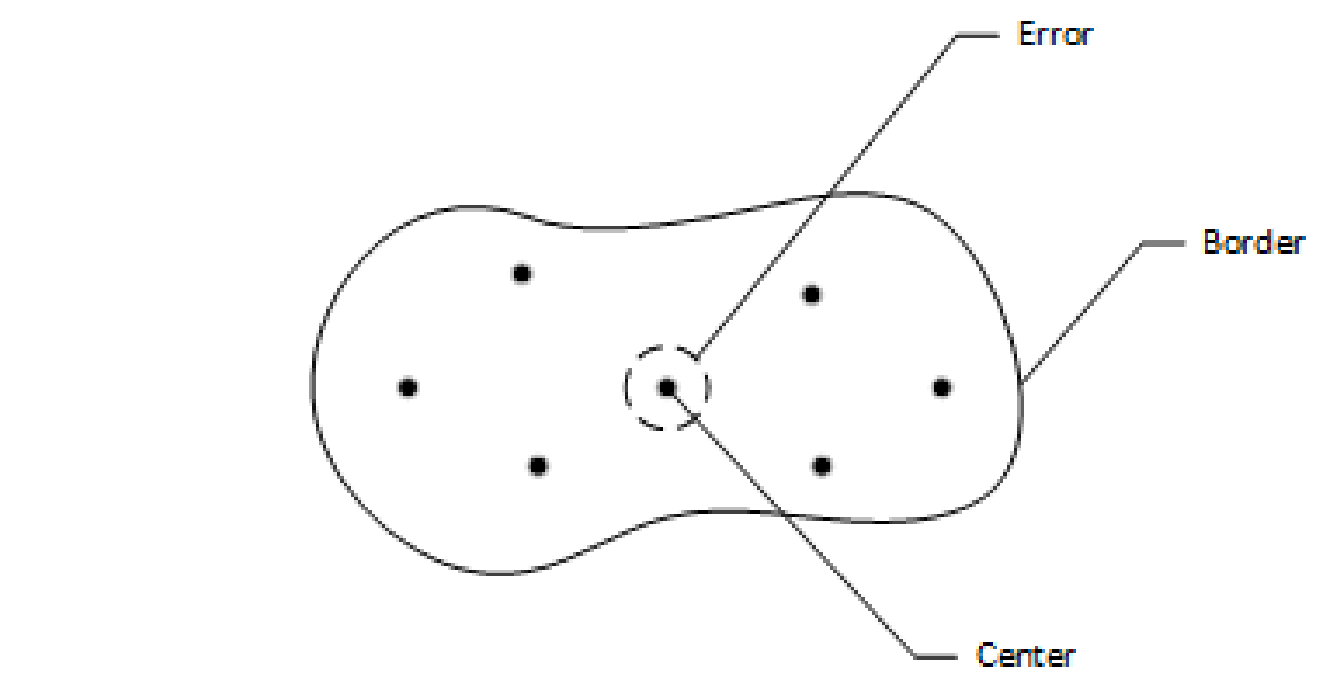}
    \caption{The parts of a group.}
    \label{group}
\end{figure}

A group, or a spatial-temporal cluster, represents several trajectories that move in similar ways. The definition of a group is based on the density of the trajectories. Density-based clustering techniques (e.g. DBSCAN) can be used to calculate it. Figure \ref{group} visually explains a group. A group contains a border, that can be fixed or variable, and a center $c_{i,t_k}$.

Formally, a group $g_i$ is a set of trajectories $t_j$, such that $g_i = \{t_1, t_2, ..., t_n\}$, where $n$ is the size of the group. Each algorithm has a different way to calculate how dense trajectories are. In general, algorithms require a maximum distance $d_{max}$ between trajectories and groups are formed from trajectories $t_j$ and $t_k$ such that $dist(t_j, t_k) \leq d_{max}$. The center of a group $c_{i,t_k}$  can be the average of the positions of all trajectories in a group at timestamp $t_k$. A radius $r^g_{error}$ is used to calculate group stops. Similar to the previous trajectory movement modeling, $c_{i, t_{i,k,stop}}$ refers to the start of the most recent (from $t_k$) series of stops.

In summary, groups also have two main types of movements, \textit{stop} and \textit{move}, as seen and formalized in Table \ref{movements}.

\begin{table}
\caption{Movement Types.}
\begin{tabu} to \linewidth {| X[2.5,c,m] | X[1.1,c,m] | X[4,c,m] |}

    \hline
    Representation & Name & Formalization \\
    \hline

    \everyrow{\taburulecolor{light-gray}\tabucline{-}}

    \raisebox{-.5\height}{\includegraphics[width=2.5cm]{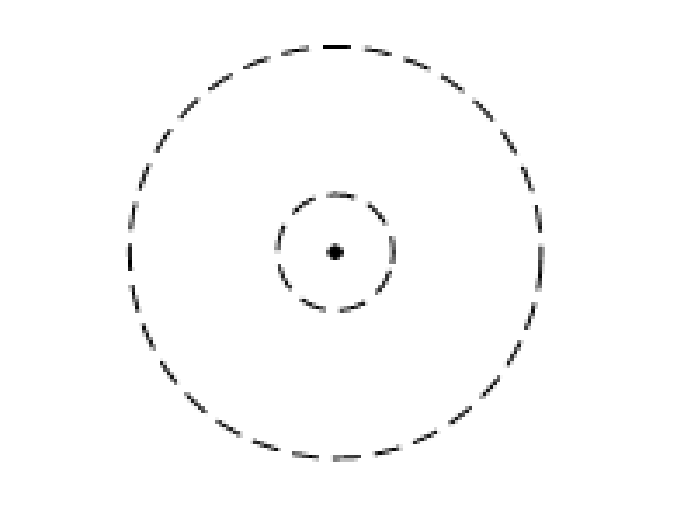}}
    & Stop & \begin{minipage}{\hsize}\begin{itemize}[label=,leftmargin=-0in] \item $movement(p_{i,t_{k-1}})=\textrm{\textit{move}}$ \item and \item $dist(p_{i, t_k}, p_{i,t_{k-1}}) \leq r_{e}$ \end{itemize}\end{minipage} \\ \taburulecolor{black}

    \raisebox{-.5\height}{\includegraphics[width=2.5cm]{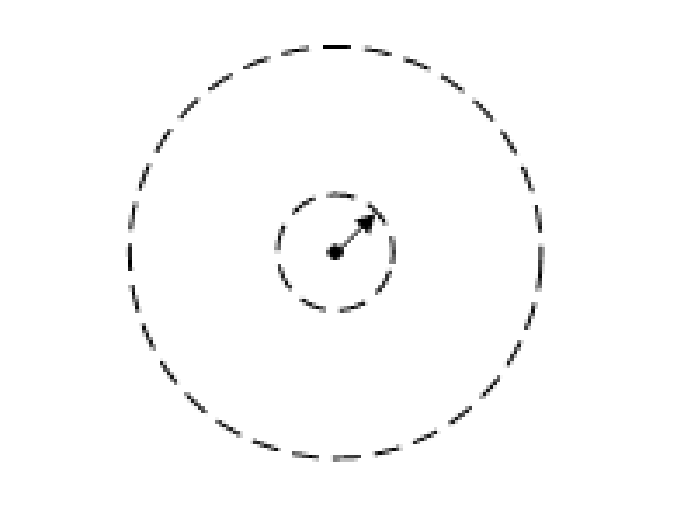}}
    & Stop & \begin{minipage}{\hsize}\begin{itemize}[label=,leftmargin=-0in] \item $movement(p_{i,t_{k-1}})=\textrm{\textit{stop}}$ \item and \item $dist(p_{i, t_k}, p_{i,t_{i,k,stop}}) \leq r_{e}$ \end{itemize}\end{minipage} \\ \taburulecolor{black}

    \raisebox{-.5\height}{\includegraphics[width=2.5cm]{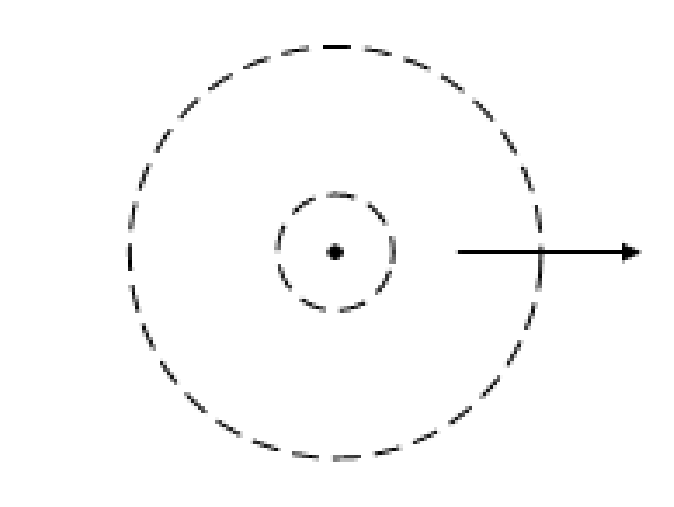}}
    & Move & \begin{minipage}{\hsize}\begin{itemize}[label=,leftmargin=-0in] \item $dist(p_{i, t_k}, p_{i, t_{i,k,stop}}) > r_n$ \item and \item $dist(p_{i, t_k}, p_{i, t_{k-1}}) > r_e$ \end{itemize}\end{minipage} \\ \taburulecolor{black}

    \raisebox{-.5\height}{\includegraphics[width=2.5cm]{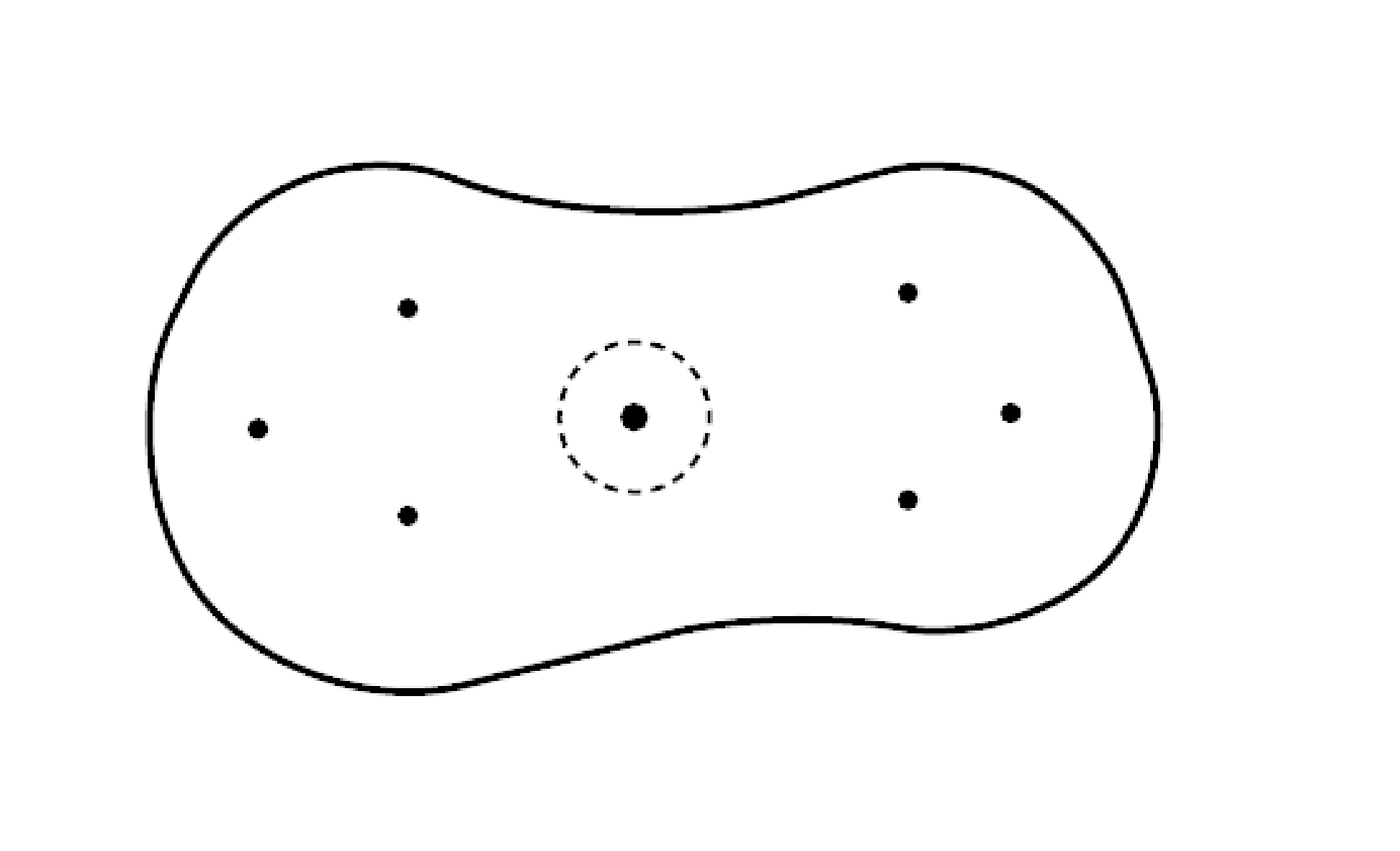}}
    & Stop & \begin{minipage}{\hsize}\begin{itemize}[label=,leftmargin=-0in] \item $dist(c_{i,t_k}, c_{i,t_{i,k,stop}}) \leq r^{g}_{error}$ \end{itemize}\end{minipage} \\ \taburulecolor{black}

    \raisebox{-.5\height}{\includegraphics[width=2.5cm]{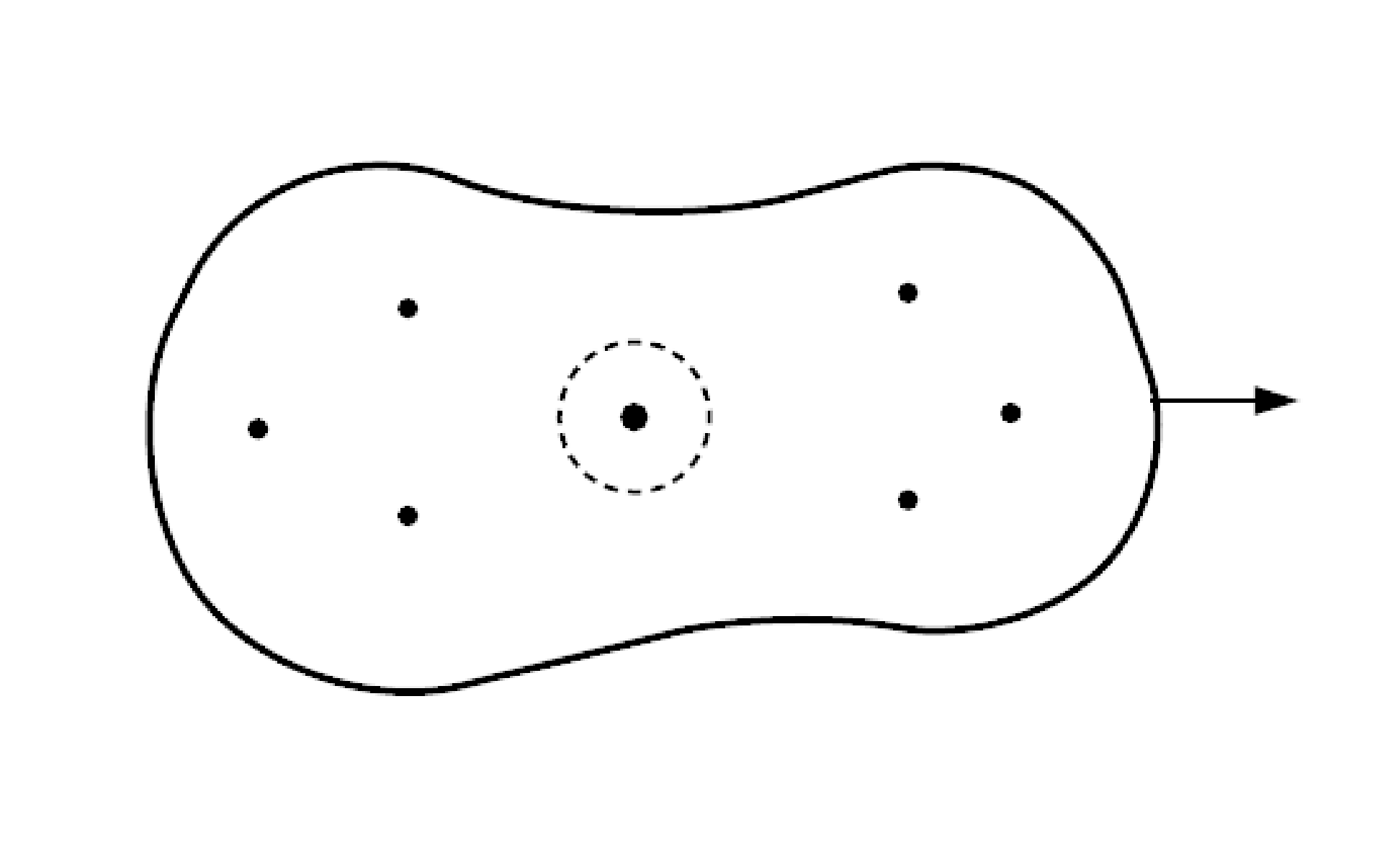}}
    & Move & \begin{minipage}{\hsize}\begin{itemize}[label=,leftmargin=-0in] \item $dist(c_{i,t_k}, c_{i,t_{i,k,stop}}) > r^{g}_{error}$ \end{itemize}\end{minipage} \\ \taburulecolor{black}

    \hline
\end{tabu}
\label{movements}
\end{table}

The formalizations previously discussed allow the definition of relations between trajectories and groups. The entire list of relations is not shown due to space restrictions. However, Table \ref{relations} presents the relations that directly guides the development of this study. The relations are mainly based on \textit{enter} or \textit{leave} movements, defined such that a group which a trajectory or a group enters to or leaves from is either stopped or moving.

\begin{table*}
\caption{Group $\times$ trajectory and group $\times$ group relations.}
\begin{tabu} to \linewidth {| X[1.9,c,m] | X[0.4,c,m] | X[3.0,c,m] | X[3.0,j,m] |}

    \hline
    Representation & Name & Formalization & Description \\
    \hline

    \everyrow{\taburulecolor{light-gray}\tabucline{-}}

    \raisebox{-.5\height}{\includegraphics[width=4.0cm]{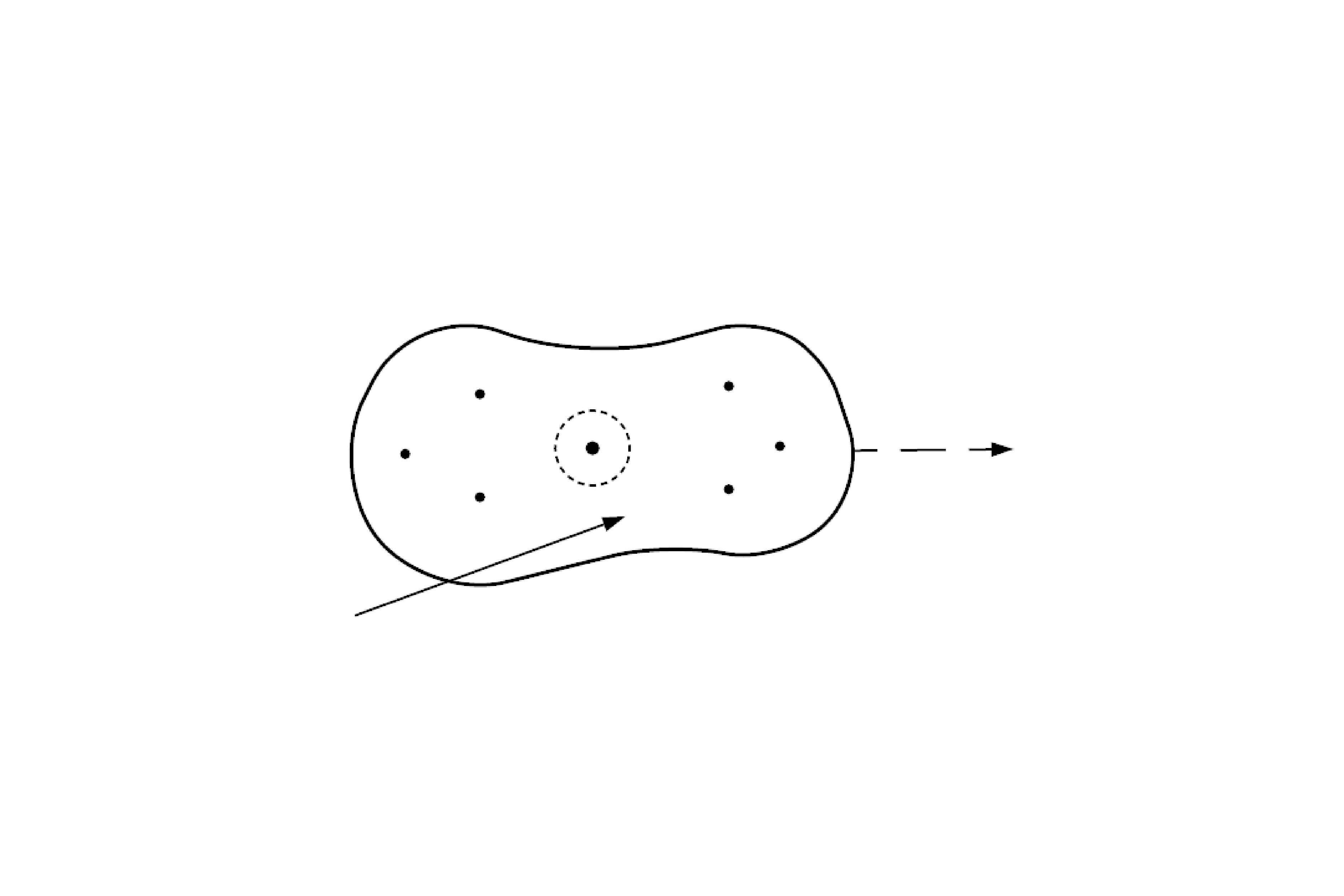}}
& Enter & \begin{minipage}{\hsize}\begin{itemize}[label=,leftmargin=-0in] \item $movement(p_{i,t_k}) = \textrm{\textit{move}}$ \item and \item $movement(c_{j,t_k}) = \textrm{\textit{stop} or \textit{move}}$ \item and \item $belong(p_{i,t_{k-1}}, group(c_{j,t_{k-1}})) = \textrm{\textit{false}}$ \item and \item $belong(p_{i,t_k}, group(c_{j,t_k})) = \textrm{\textit{true}}$ \end{itemize}\end{minipage} &
    \begin{minipage}{\hsize}Represents a relation between a group and a trajectory in which the trajectory enters the cluster, according to the clustering algorithm. The dotted line to the right of the cluster indicates that the cluster may or may not be moving.\end{minipage} \\ \taburulecolor{black}

    \raisebox{-.5\height}{\includegraphics[width=4.0cm]{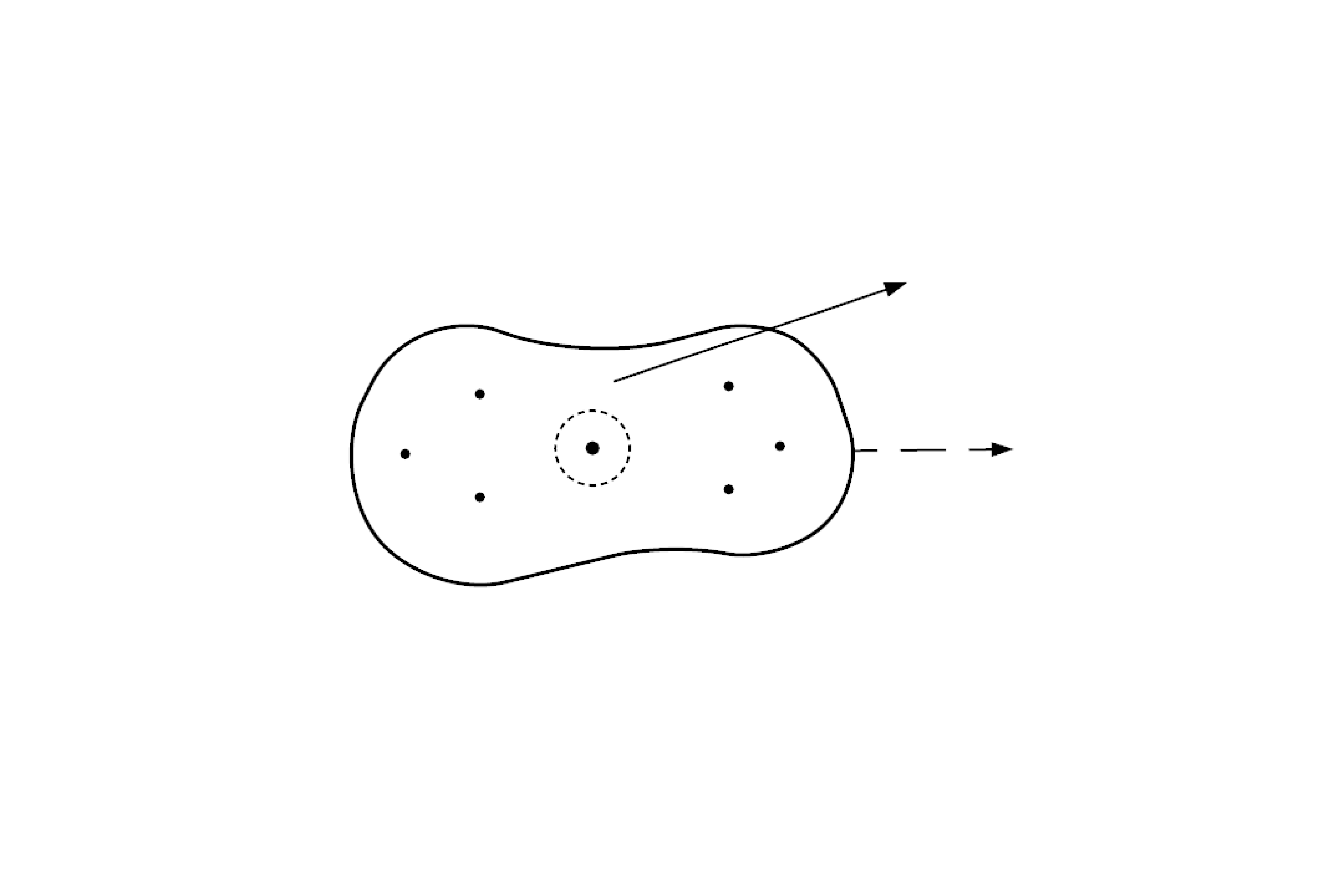}}
& Leave & \begin{minipage}{\hsize}\begin{itemize}[label=,leftmargin=-0in] \item $movement(p_{i,t_k}) = \textrm{\textit{move}}$ \item and \item $movement(c_{j,t_k}) = \textrm{\textit{stop} or \textit{move}}$ \item and \item $belong(p_{i,t_{k-1}}, group(c_{j,t_{k-1}})) = \textrm{\textit{true}}$ \item and \item $belong(p_{i,t_k}, group(c_{j,t_k})) = \textrm{\textit{false}}$ \end{itemize}\end{minipage} &
    \begin{minipage}{\hsize}Represents a relation between a group and a trajectory in which the trajectory leaves the cluster, according to the clustering algorithm. The dotted line to the right of the cluster indicates that the cluster may or may not be moving.\end{minipage} \\ \taburulecolor{black}



    \raisebox{-.5\height}{\includegraphics[width=4.0cm]{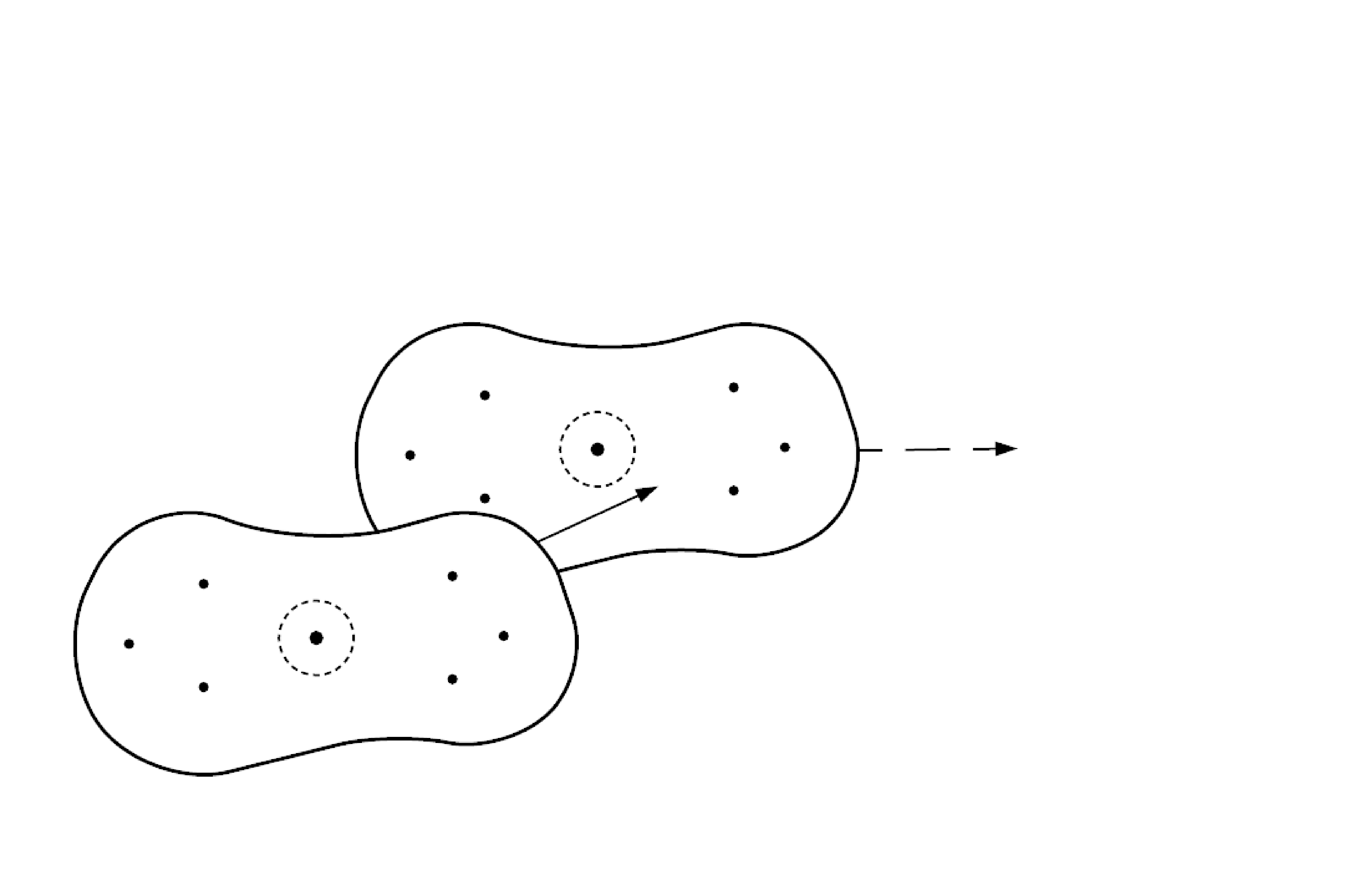}}
    & Merge & \begin{minipage}{\hsize}\begin{itemize}[label=,leftmargin=-0in] \item $movement(c_{i,t_k}) = \textrm{\textit{stop}}$ \item and \item $movement(c_{j,t_k}) = \textrm{\textit{stop} or \textit{move}}$ \item and \item $belong(group(c_{i,t_{k-1}}), group(c_{j,t_{k-1}})) = \textrm{\textit{false}}$ \item and \item $belong(group(c_{i,t_k}), group(c_{j,t_k})) = \textrm{\textit{true}}$ \end{itemize}\end{minipage} &
    \begin{minipage}{\hsize}Represents a relation between two clusters, in which a cluster enters, or merges with, another cluster. Usually, the two clusters are not distinguishable after the merge as clustering algorithms show only one resulting cluster. The dotted line to the right of the cluster indicates that the cluster may or may not be moving.\end{minipage} \\ \taburulecolor{black}

    \raisebox{-.5\height}{\includegraphics[width=4.0cm]{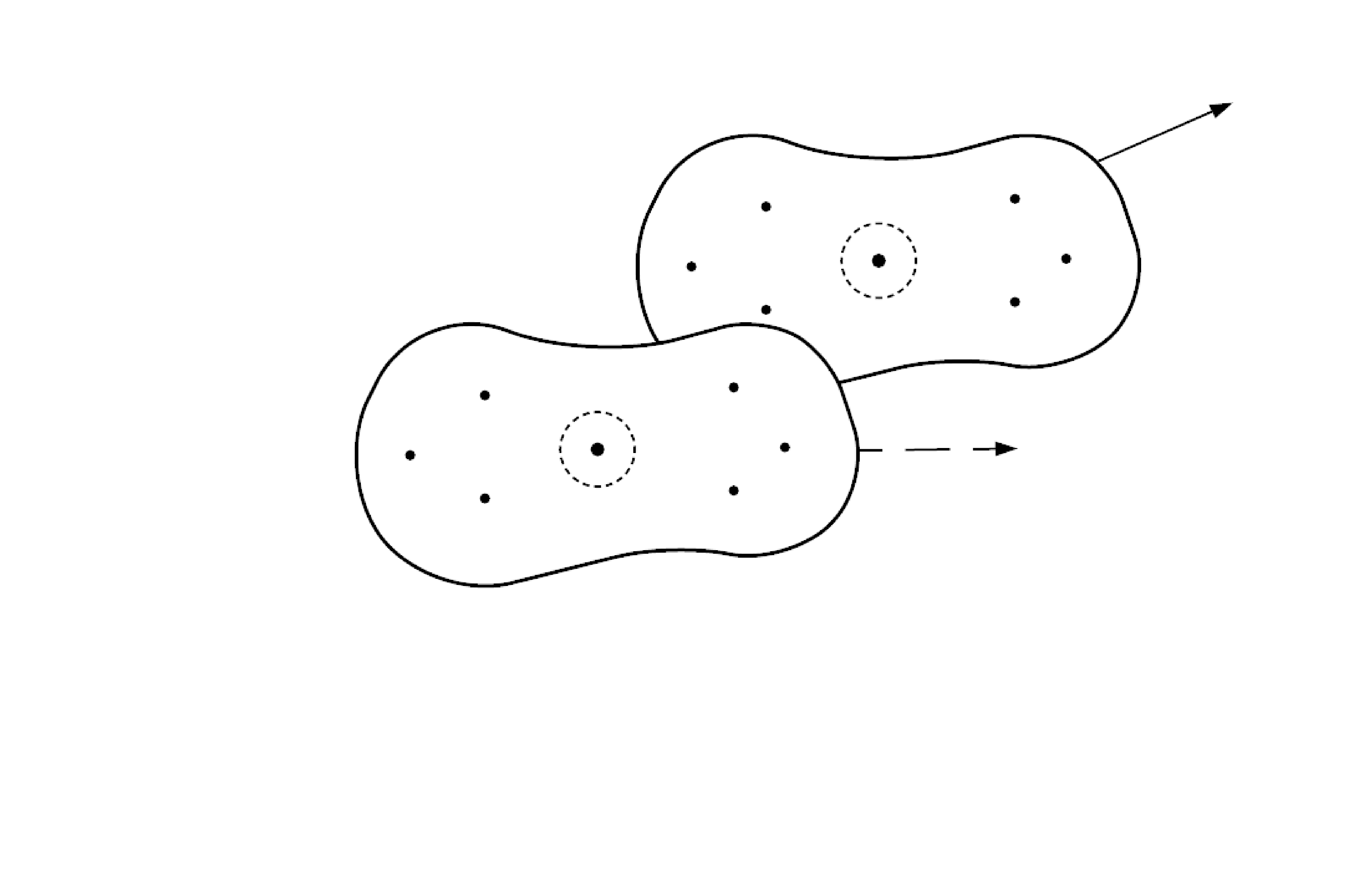}}
    & Split & \begin{minipage}{\hsize}\begin{itemize}[label=,leftmargin=-0in] \item $movement(c_{i,t_k}) = \textrm{\textit{move}}$ \item and \item $movement(c_{j,t_k}) = \textrm{\textit{stop} or \textit{move}}$ \item and \item $belong(group(c_{i,t_{k-1}}), group(c_{j,t_{k-1}})) = \textrm{\textit{true}}$ \item and \item $belong(group(c_{i,t_k}), group(c_{j,t_k})) = \textrm{\textit{false}}$ \end{itemize}\end{minipage} &
    \begin{minipage}{\hsize}Represents a relation between two clusters, in which a cluster leaves, or splits from, another cluster. Usually the two clusters are not distinguishable before the split as clustering algorithms show only one original cluster. The dotted line to the right of the cluster indicates that the cluster may or may not be moving.\end{minipage} \\ \taburulecolor{black}



    \hline
\end{tabu}
\label{relations}
\end{table*}

Clusters appear and disappear. During their lifetime, many of the previously discussed relations may happen multiple times. These relations can be listed in order of occurrence, accross many timestamps, for a single cluster. This can be done even if the cluster slowly replaces all of its original moving elements. This list is the cluster lifecycle. The lifecycle of a cluster is described in terms of the relations the cluster has with trajectories and other clusters.

Similar relationships for cluster lifecycle analysis are defined based on the discussion in this technical report and on \cite{Li20132752}, and are presented in Table \ref{cluster_relations}.

\begin{table}

\caption{Cluster Relations.}
\begin{tabular}{|m{3cm}|m{0.7cm}|m{3.8cm}|}
\hline
Representation & Name  & Description                                                                      \\
\hline
\includegraphics[width=3cm]{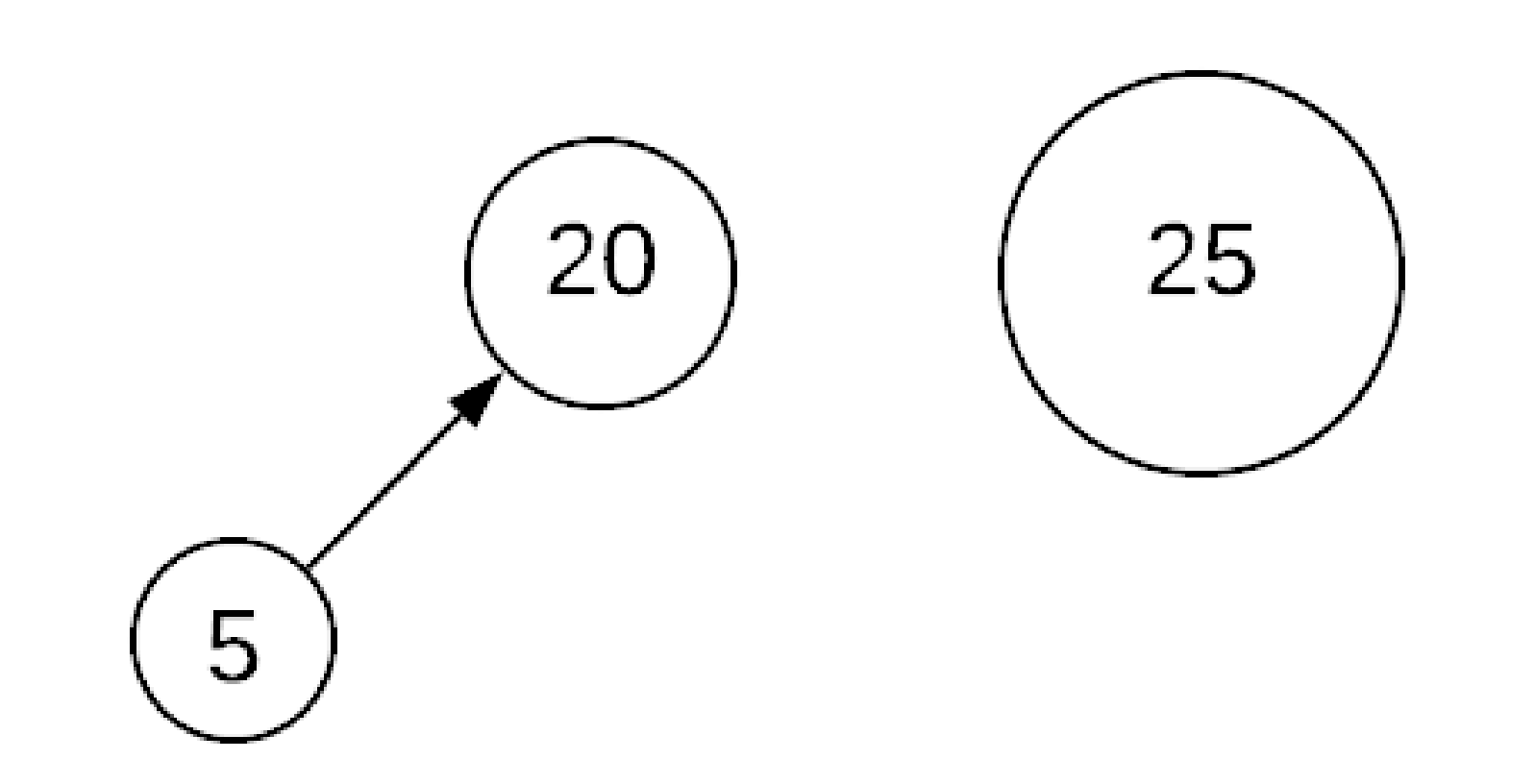}
& Enter & Describes the moment in which individual trajectories or an entire cluster enters the cluster.  \\ \hline
\includegraphics[width=3cm]{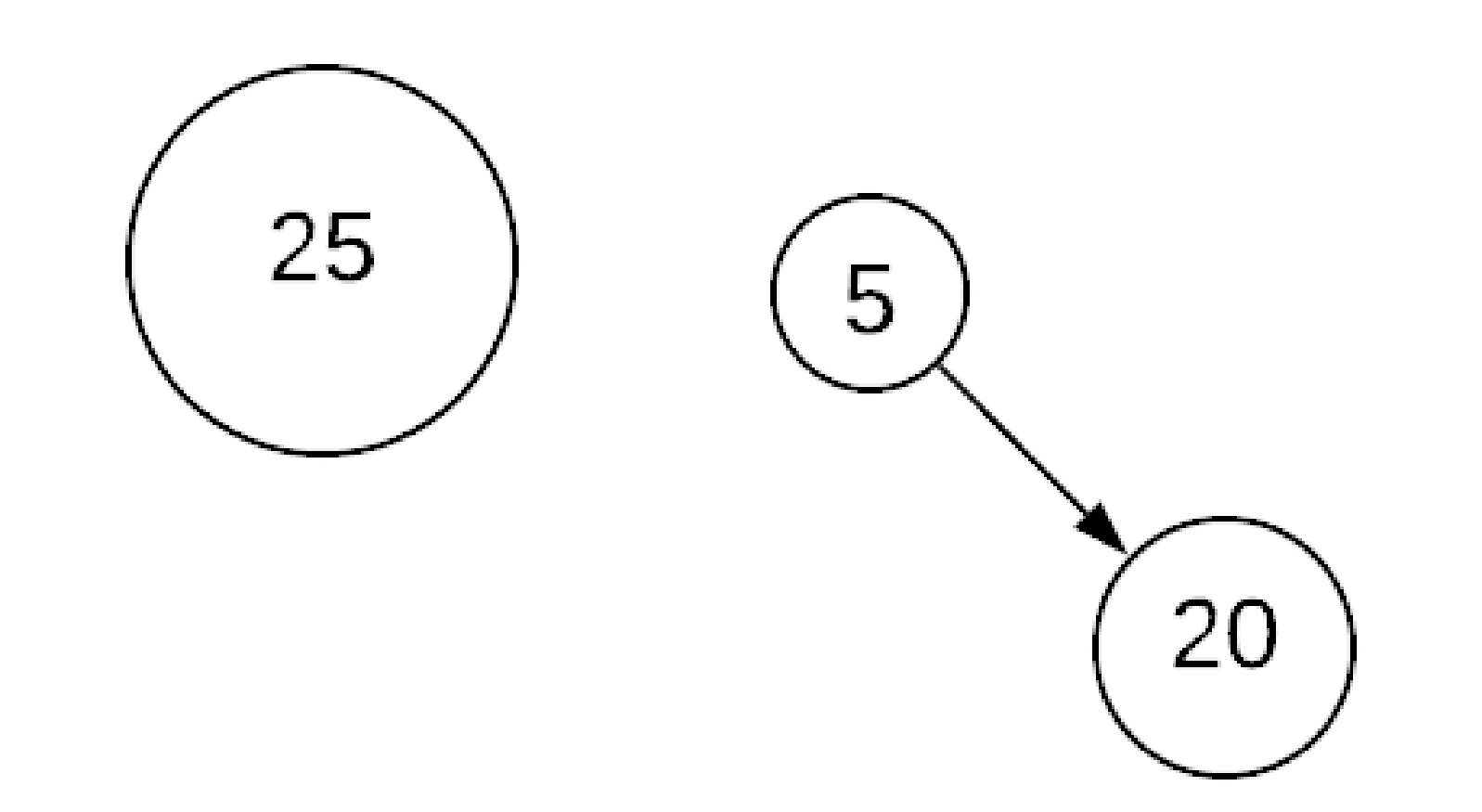}
& Leave & Describes the moment in which independent trajectories or an entire cluster leaves the cluster. \\ \hline
\includegraphics[width=2.5cm]{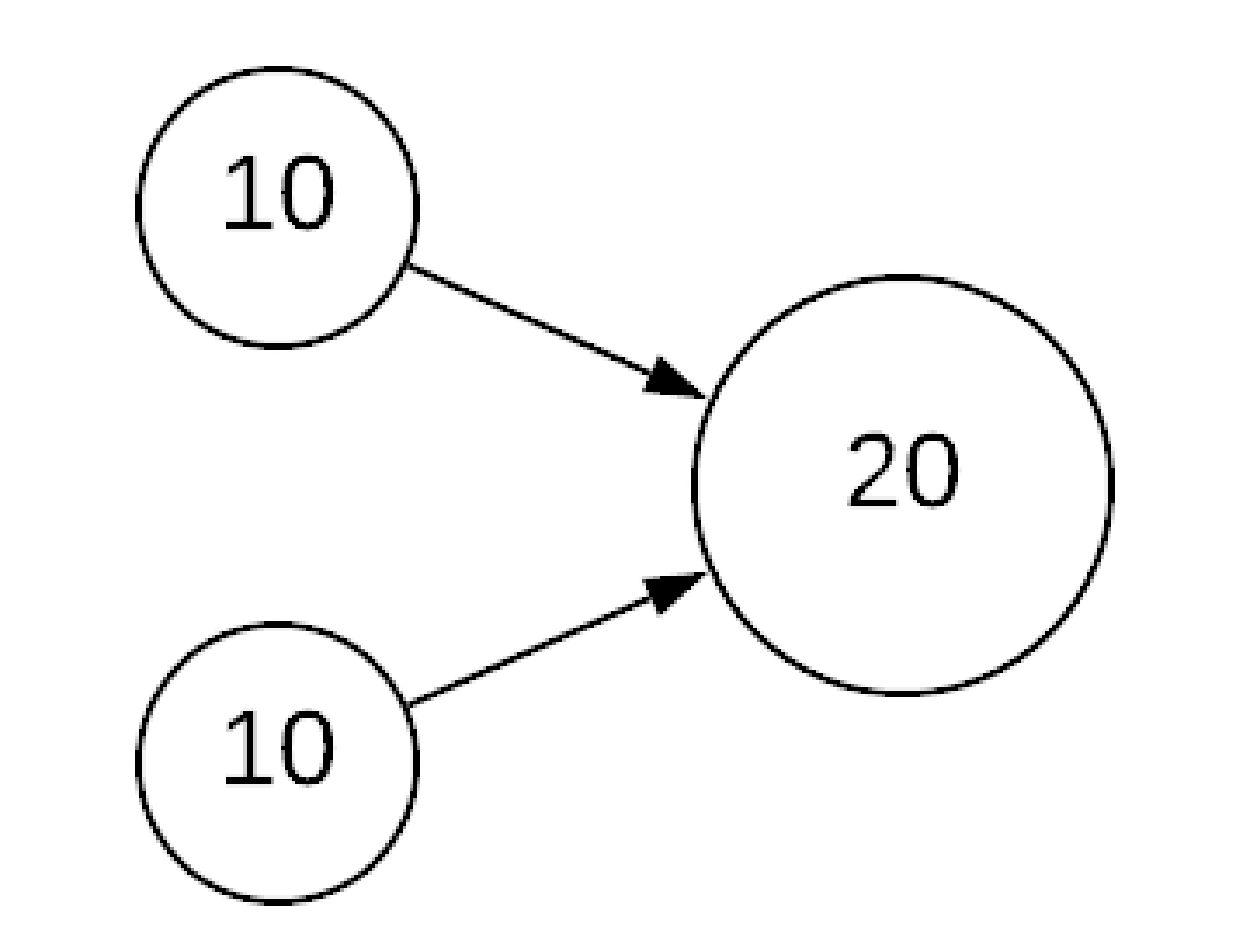}
& Merge & Describes the moment in which two or more clusters combine to form a new cluster.               \\ \hline
\includegraphics[width=2.5cm]{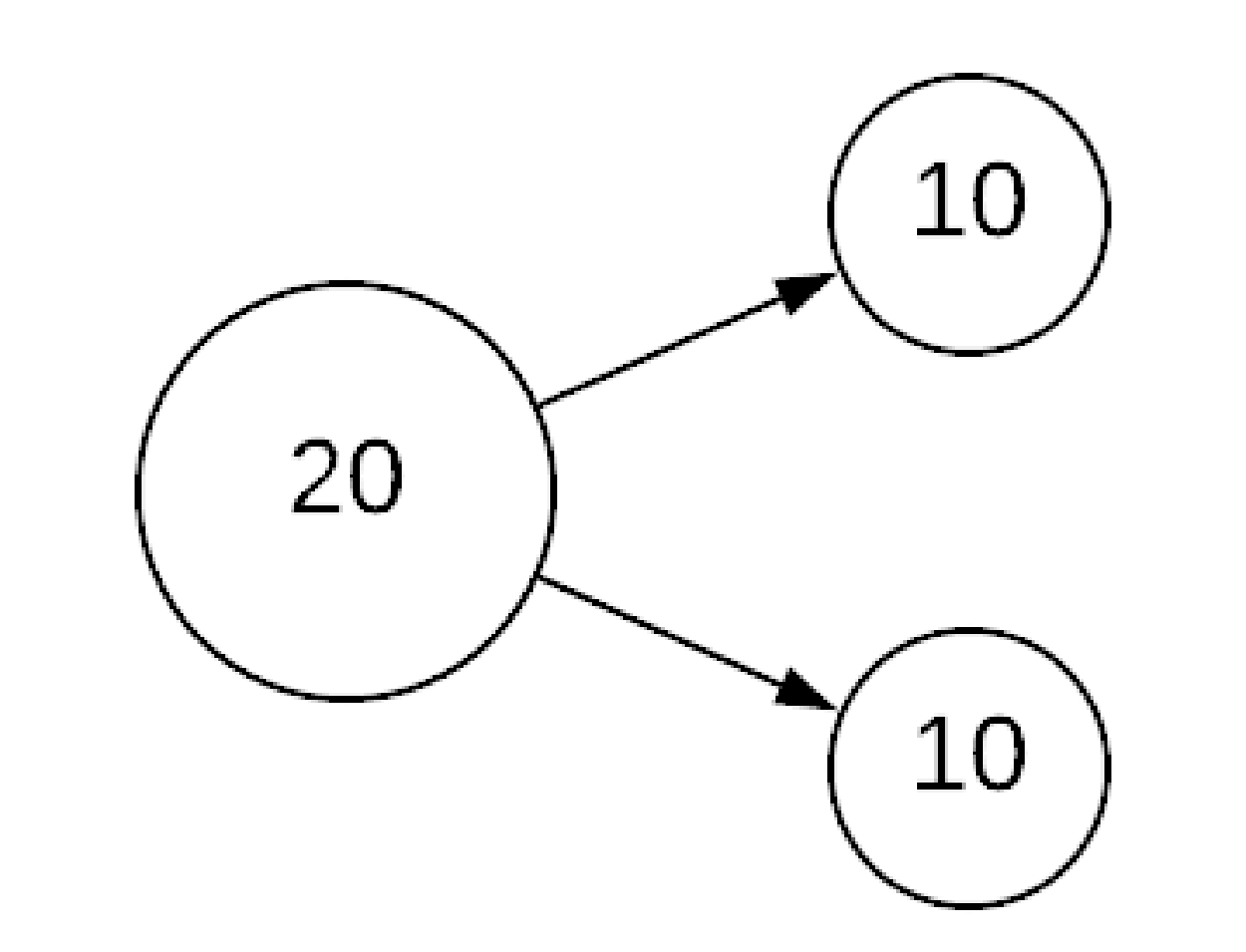}
& Split & Describes the moment in which a cluster is divided in two or more clusters.                     \\ \hline
\includegraphics[width=2.5cm]{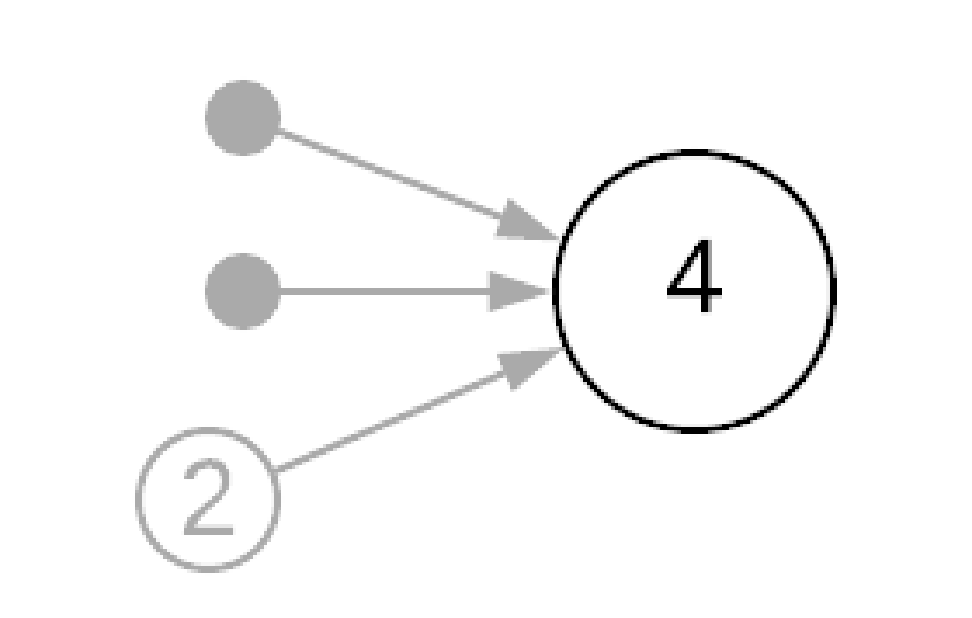}
& Start & Marks the beginning of the existence of a cluster.                                              \\ \hline
\includegraphics[width=2.5cm]{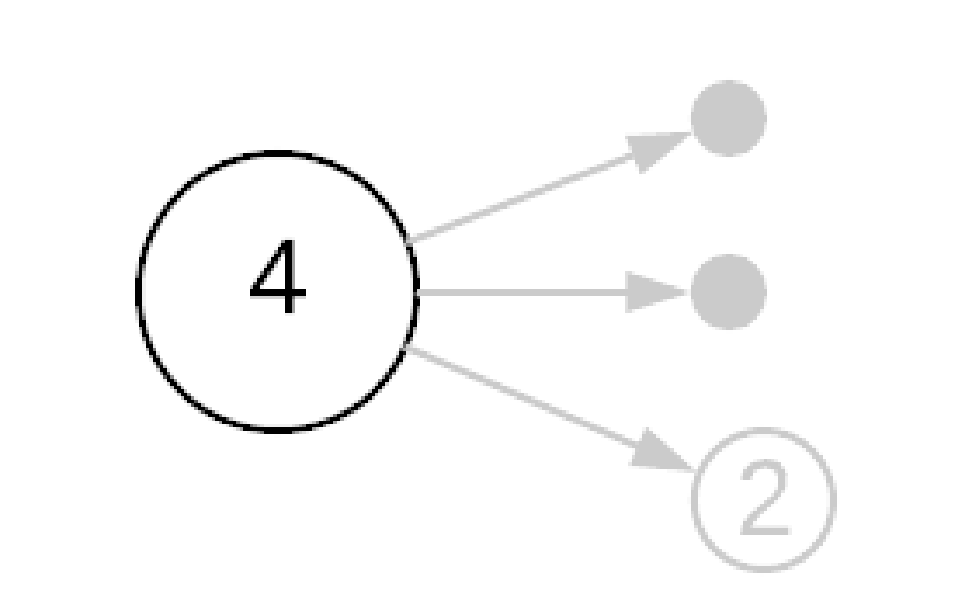}
& End   & Marks the expiration of a cluster.                                                              \\ \hline
\end{tabular}
\label{cluster_relations}
\end{table}

Some important definitions are required before analyzing the lifecycle of a cluster. First, a valid cluster is one that has a minimum number of elements, \(minCluster\). For example, three vehicles together may or may not form a cluster.

Second, the similarity between clusters in two timestamps depends on the percentage of shared elements, \(minShared\). Consider a cluster formed with 10 elements. At the next timestamp, the algorithm identifies another cluster, with 8 elements, all of them from the set of the 10 elements of the previous timestamp. Is the second cluster the same as first? What if the second cluster had 14 elements, but only 4 of them were a subset of the initial 10 elements? This similarity is formalized in equation (\ref{similarity}).

\begin{equation}
\label{similarity}
\begin{split}
c_1 = c_2 \iff nu&m\_elem(intersection(c_1, c_2)) > \\ &minShared * num\_elem(c_1) \enskip \land \\ nu&m\_elem(intersection(c_1, c_2)) > \\ &minShared * num\_elem(c_2)
\end{split}
\end{equation}

where \(c_1\) and \(c_2\) are clusters, \(num\_elem(c)\) calculates the number of elements of a cluster, \(intersection(c_1, c_2)\) returns the number of elements that are present in clusters \(c_1\) and \(c_2\). Informally, it means that if more than \(minShared\) percent of elements of the first cluster is present on the second cluster and more than \(minShared\) percent elements of the second cluster is present on the first cluster, then the two clusters are the same.

Third, to reduce the number of cluster comparisons, only clusters whose centroids are near are compared. The maximum distance of centroids is defined as \(maxDistCentroid\).

The algorithm creates two dictionaries with the clusters at the previous and the current timestamps as the keys. For each cluster \(c_i\) of the first timestamp, the algorithm compares the distance between \(c_i\) and the other clusters \(c_j\) in the timestamp. If the distance of their centroids is less than \(maxDistCentroid\), then \(c_j\) is added to the dictionary as a value under the key \(c_i\). This process is repeated for clusters in the second timestamp. The first dictionary identifies leave, split, or end relationships, whereas the second one identifies enter, merge, or start relationships.

The algorithm then checks the number of shared elements between each key-value pair. The algorithm then assigns a label to the values indicating whether the two clusters (key and value) are the same, and another label for a potential leave (if processing the first dictionary) or enter (if processing the second dictionary) relationships. Once all values of a key are processed, the key (or the cluster associated with the key) receives a final label based on the number of leave or enter relationships.

\subsection{Big Data Framework}

Spatial-temporal data, such as trajectories, are rich in volume and value. When analyzing this type of data, big data approaches are needed to reach conclusions in a reasonable amount of time. This study identified some opportunities for parallel processing and proposes a big data framework for cluster lifecycle analysis.

\begin{figure}
    \centering
    \includegraphics[width=\linewidth]{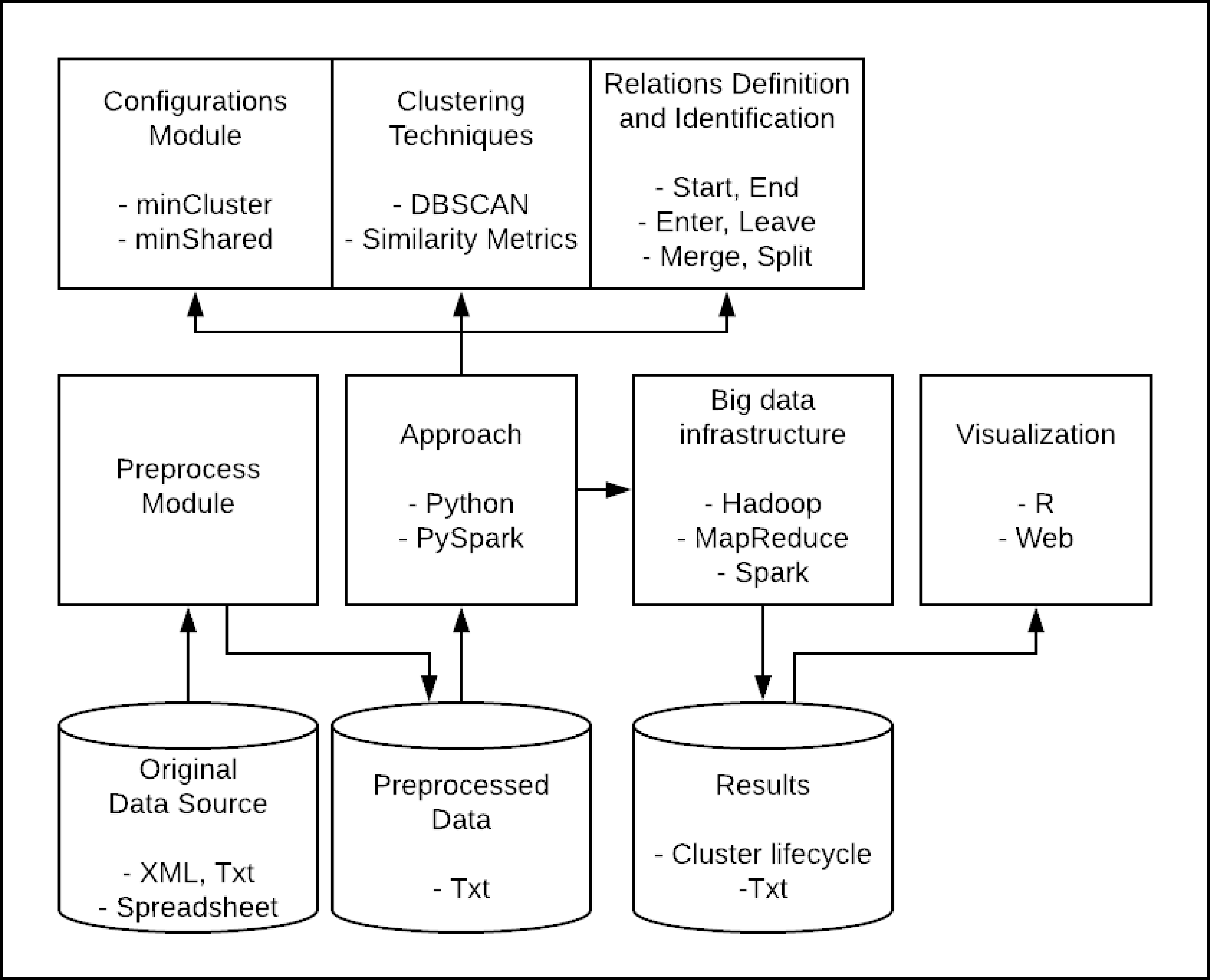}
    \caption{The cluster lifecycle analysis framework.}
    \label{big_data_framework}
\end{figure}

First, data can be divided into groups based on where they were measured. This is done by dividing the world map into a grid, and performing parallel calculations on each grid section. The drawback is that clusters that cross a grid border may have their analyses affected. Second, the dictionary for each timestamp can be processed in parallel by multiple machines since they are independent. This parallel computation improves the processing time and could produce significant results when expanded to massive amounts of data. Figure \ref{big_data_framework} shows a big data framework for cluster lifecycle analysis.

\section{Conclusions and Future Work}
\label{conclusion}

The study of trajectory clustering lacks approaches that analyze clusters as a whole, from a cluster creation to its disappearance, including what happens during the cluster lifetime. This study proposes an ongoing approach that analyzes individual trajectories, identifies clusters and their relationships with other trajectories or clusters, and stores these relationships to form a cluster lifecycle.

The main contributions of this study are \begin{enumerate*}[label=(\roman*)] \item a list of relations that govern the lifecycle of a cluster, \item an algorithm to identify those relations, and \item a big data framework to support cluster lifecycle analysis \end{enumerate*}.

In the future, more analyses are required on the lifecycle of clusters to identify trends and allow predictions about their creation, behavior, or disappearance. Additionally, it is worth investigating and discovering hierarchical relationships between clusters (e.g. nested clusters), since identifying new relationships may lead to new forms of analysis. Lastly, more research on the distributed clustering approach can be performed to impact algorithm performance.

\section*{Acknowledgment}

The authors would like to thank the Natural Sciences and Engineering Research Council of Canada (NSERC) and CANARIE for the financial support.

\bibliographystyle{IEEEtran}
\bibliography{./bibliography/scopus,./bibliography/noscopus}

\end{document}